\documentclass[prl,twocolumn,twoside,preprintnumbers,superscriptaddress,letterpaper]{revtex4}
\usepackage[utf8]{inputenc}
\usepackage{setup}
\usepackage{soul} 

\preprint{NIKHEF 2020-034}
\preprint{ZU-TH 34/20}
\preprint{IPPP/20/45}
\preprint{CERN-TH-2020-163}

\begin{document}

\title{Precise predictions for $\PW\PH$+jet production at the LHC}

\author{R.~Gauld}
\email{r.gauld@nikhef.nl}
\affiliation{Nikhef, Science Park 105, NL-1098 XG Amsterdam, The Netherlands}

\author{A.~\surname{Gehrmann--De~Ridder}}
\email{gehra@phys.ethz.ch}
\affiliation{Institute for Theoretical Physics, ETH, CH-8093 Zürich, Switzerland}
\affiliation{Department of Physics, University of Zürich, CH-8057 Zürich, Switzerland}

\author{E.~W.~N.~Glover}
\email{e.w.n.glover@durham.ac.uk}
\affiliation{Institute for Particle Physics Phenomenology, University of Durham, DH1 3LE Durham, United Kingdom}

\author{A.~Huss}
\email{alexander.huss@cern.ch}
\affiliation{Institute for Particle Physics Phenomenology, University of Durham, DH1 3LE Durham, United Kingdom}
\affiliation{Theoretical Physics Department, CERN, CH-1211 Geneva 23, Switzerland}

\author{I.~Majer}
\email{majeri@phys.ethz.ch}
\affiliation{Institute for Theoretical Physics, ETH, CH-8093 Zürich, Switzerland}

\date{\today}

\begin{abstract}
We present precise predictions for the production of a Higgs boson in association with a hadronic jet and a $\PW$ boson at hadron colliders.
The behaviour of QCD corrections are studied for fiducial cross sections and distributions of the charged gauge boson and jet-related observables.
The inclusive process (at least one resolved jet) and the exclusive process (exactly one resolved jet) are contrasted and discussed.
%
The inclusion of QCD corrections up to $\order*{\alphas^3}$ leads to a clear stabilisation of the predictions and contributes substantially to a reduction of remaining theoretical uncertainties.
\end{abstract}

\maketitle

\section{Introduction}

The production of a Higgs boson \PH in association with either a \PWpm or \PZ boson and possible hadronic jets, also known as the Higgs Strahlung process, is one of the most promising Higgs production modes for the accurate determination of the Higgs-boson couplings with known Standard Model particles.
These classes of Higgs production channels provide the opportunity to probe the gauge-boson--Higgs vertex ($VV\PH$) separately for $V=\PWpm$ and $V=\PZ$ and give access to the decay of the Higgs boson into a quark--antiquark pair (e.g. bottom or charm).
The presence of a leptonically decaying vector boson provides a clean experimental signature, enhancing the control over the backgrounds of the $V\PH$ process when hadronic Higgs boson decays are considered.
In 2017, the LHC experiments~\cite{Aaboud:2018zhk,Sirunyan:2018kst} announced the observation of a Standard Model Higgs-like particle decaying to a pair of bottom--antibottom quarks, precisely through this Higgs Strahlung production channel.

First differential measurements based on simplified template cross sections have been reported in ref.~\cite{Aaboud:2019nan}. Updated measurements with the full Run 2 data set were presented in~refs.~\cite{Aad:2020jym,Aad:2020eiv}, indicating that the observed production rate is consistent with the expectation of the Standard Model within experimental uncertainties of $\approx 20\%$. These (differential) measurements are currently limited by the available statistics, but will eventually become systematically limited. For example, the uncertainty on the extracted production rate due to signal modelling constitutes $\approx (5-8)\%$ of the total uncertainty~\cite{Aad:2020jym}.

As both the measurement and interpretation of this data relies on theoretical knowledge of the $V\PH$(+jet) production and decay modes within the Standard Model, it is of critical importance to have more precise theoretical predictions both for fiducial cross sections and for differential distributions in the kinematic regions probed by experiments.
This includes the study of QCD effects related to Higgs production and decay.
An improved understanding of these effects will be of particular relevance for future measurements of $V\PH$(+jet) production---including data from Run~III of the LHC and the planned high-luminosity upgrade (HL-LHC).

%


Depending on the selected $V\PH$(+jet) production mode, different theoretical knowledge is currently available.
For $V\PH$ production, the total inclusive cross section has been known to NNLO QCD precision for some time, and theoretical predictions are available publicly through the program \texttt{VH@NNLO}~\cite{Brein:2012ne}.
Fully differential NNLO predictions for $V\PH$ observables obtained via the combination of Higgs production and its decay to bottom--antibottom quarks have been presented in refs.~\cite{Ferrera:2017zex,Caola:2017xuq,Gauld:2019yng}.
These three computations have one essential feature in common: they consider massless \Pqb quarks except in the bottom Yukawa coupling and use the flavour-$k_t$ algorithm~\cite{Banfi:2006hf} to define \Pqb-jets.
In the case of $\PW\PH$ production, a computation using massive \Pqb quarks and standard jet algorithms has been recently reported in ref.~\cite{Behring:2020uzq}.
The theoretical predictions for $V\PH$+jet are currently limited to NLO accuracy and additionally consider the Higgs boson as stable.
The NLO QCD computations for $V\PH$ and $V\PH$+jet production have been merged in the context of parton showers to provide full NLO+PS simulations for these (on-shell) Higgs production modes in refs.~\cite{Hamilton:2012rf,Luisoni:2013kna,Astill:2018ivh}, which have been further refined to include electroweak corrections in~\cite{Granata:2017iod}.

It is the purpose of this paper to provide a new level of theoretical precision for observables related to the $V\PH$+jet production mode
by including QCD corrections to the Drell--Yan-type and top quark loop-induced contributions to this production mode up to orders $\alphas^3$ and $\alphas^2y_t$, respectively.
This particular mode is of phenomenological interest as the requirement of a resolved jet has the potential to provide more differential information on the production process and sensitivity to QCD radiation effects.
An improved theoretical understanding of both inclusive (requiring at least one resolved jet) and exclusive (requiring exactly one resolved jet) $V\PH$+jet production is also of critical importance to experimental analyses which rely on the use of exclusive jet bins, as highlighted in refs.~\cite{Aad:2020jym,Aad:2020eiv}.

We will specifically consider $\PWp\PH$+jet production where the charged vector boson decays leptonically and the Higgs boson is produced on-shell. This can be regarded as a first step towards computing $V\PH$+jet observables including the decay of the Higgs boson into a quark--antiquark pair, with high theoretical precision.





In the following, we provide details about the ingredients of the computation before presenting results for \SI{13}{\TeV} LHC runs in terms of fiducial cross sections and a selection of differential distributions. 

\section{Details of the Calculation}
Throughout this work, we compute observables related to the production of a Higgs boson \PH, together with a positively charged weak gauge boson \PWp, and an additional hadronic jet including up to $\order*{\alphas^3}$ corrections in perturbative QCD.
That is, we consider the process
\begin{equation}
    \Pp \Pp \to  \PH\PWp + \text{jet} \to \PH + \Plp \Pnulepton + \text{jet,}
\end{equation}
where the Higgs boson is produced on-shell and the charged vector boson decays leptonically (including all spin-correlation and off-shell effects).

The calculation of all contributions is carried out within the \nnlojet framework~\cite{Ridder:2015dxa}: a fixed-order parton-level event generator using the NNLO antenna subtraction formalism~\cite{GehrmannDeRidder:2005cm,GehrmannDeRidder:2005aw,GehrmannDeRidder:2005hi,Daleo:2006xa,Daleo:2009yj,Boughezal:2010mc,Gehrmann:2011wi,GehrmannDeRidder:2012ja,Currie:2013vh} to regulate infrared divergences that appear in different partonic processes beyond leading order.
The LO contribution to this process begins at $\order*{\alphas}$.
Starting from $\order*{\alphas^2}$, two types of contributions, which are commonly referred to as Drell--Yan-type and heavy quark-loop induced, can be distinguished.

The Drell--Yan-type contributions arise from contributions akin to the process of $\PW$+jet production, where the Higgs boson is emitted from the intermediate gauge-boson propagator.
As such, the corresponding predictions can be obtained using essential components of the $\PW$+jet calculation~\cite{Gehrmann-DeRidder:2019avi} already available in \nnlojet.
The necessary amplitudes for the $\PW\PH$+jet partonic processes were constructed from those of the $\PW$+jet case by inserting a \PW--\PH vertex onto the $\PW$ propagator.
A subset of the new $\order*{\alphas^3}$ Drell--Yan-like contributions were independently derived in addition to amplitudes provided by the OpenLoops~2~\cite{Buccioni:2019sur} library.
As a consequence of their shared QCD structure, the subtraction terms of the Drell--Yan-type contributions can be readily constructed from the $\PW$+jet subtraction terms computed at the same order.

The heavy quark loop-induced contributions to $\PW\PH$+jet production begin at $\order*{\alphas^2y_t}$, i.e.~at NLO level.
We only consider the dominant contributions enhanced by the top Yukawa coupling, that is where the Higgs boson couples to a top-quark loop.
The corresponding loop amplitudes are known in the literature as so-called $R_I$-type matrix-elements and are related to the Higgs Strahlung production process without the additional jet requirement~\cite{Brein:2011vx}.
In the present computation, top-loop induced contributions to the cross section arise only through the interference of such one-loop amplitudes with Drell--Yan-type amplitudes
and are therefore of $\mathcal{O}(\alpha_s^2 y_t)$.
Contributions of higher order in $\alpha_s$ or $y_t$ are mostly unknown and not included here.
As will be highlighted in what follows, such higher-order terms are not expected to be of phenomenological relevance.
%

\section{Numerical Results at 13 TeV}

\subsection{Numerical Set-up}

We present predictions for $\sqrt{s} = \SI{13}{\TeV}$ proton--proton collisions using the parton distribution function set \texttt{NNPDF31\_nnlo\_as\_0118} from the LHAPDF library~\cite{Buckley:2014ana}.
We require a hard cut of $p_\bot > \SI{20}{\GeV}$ for each identified final-state jet, which are clustered with the anti-$k_t$ algorithm using $\Delta R = 0.5$.
We demand at least one resolved jet to be present in the final state, which defines the \emph{inclusive} production process $\sigma_{\geq1\text{j}}$.
In addition, we also consider the \emph{exclusive} process, denoted $\sigma_{1\text{j}}$, where additional resolved jets are vetoed.
The charged leptons are subject to a transverse momentum cut of $p_{\bot,\Pl} > \SI{25}{\GeV}$ and a cut $\abs{y_{\Pl}} < 2.5$ in the absolute value of their rapidity.
Lastly the missing transverse energy must satisfy~$E_{\bot,\text{miss}} > \SI{25}{\GeV}$.

In the following, we collect the values of all independent parameters used in the computation (based on the $G_\mu$ electroweak scheme):
The \PW-boson mass and width $M_{\PW} = \SI{81.385}{\GeV}$, $\Gamma_{\PW} = \SI{2.085}{\GeV}$, the \PZ-boson mass $M_{\PZ} = \SI{91.1876}{\GeV}$, the Fermi constant $G_\text{F} = \SI{1.1663787e-5}{\GeV^{-2}}$, and the top-quark pole mass $m_{\Pqt} = \SI{173.21}{\GeV}$.
In addition, the theoretical predictions are obtained with a diagonal CKM matrix.

In order to estimate the theoretical uncertainty of the inclusive predictions we vary the factorisation and renormalisation scales by a factor of two around the central value of the dynamical mass of the $\PW\PH$ system according to the commonly used 7-point variation scheme:
\begin{equation*}
	\muF = M_{\PW\PH} \, \qty[ 1, \tfrac{1}{2}, 2], \qquad
	\muR = M_{\PW\PH} \, \qty[ 1, \tfrac{1}{2}, 2],
\end{equation*}
with the constraint $\tfrac{1}{2}\leq \muF/\muR \leq 2$.
The analytic dependence on the renormalisation scale has been explicitly verified following \cite{Currie:2018xkj}.
In the case of the exclusive process, theoretical uncertainties can be underestimated by such a correlated scale variation as discussed in Ref.~\cite{Stewart:2011cf}.
We therefore adopt the more conservative uncorrelated prescription introduced in~\cite{Stewart:2011cf}
\begin{align}
    \label{eq:exerr}
    \sigma_{1\text{j}} &\equiv \sigma_{\geq1\text{j}} - \sigma_{\geq2\text{j}} , &
    \Delta^2_{1\text{j}} &= \Delta^2_{\geq1\text{j}} + \Delta^2_{\geq2\text{j}} ,
\end{align}
where $\Delta_{\geq1(2)\text{j}}$ denote the uncertainties for inclusive $\PWp\PH$+1(2)-jet production obtained from their respective 7-point scale variation.

\subsection{Fiducial Cross Sections}

The fiducial cross sections for inclusive (``$\geq 1$ jet'') and exclusive (``1 jet'') $\PWp\PH$+jet production are summarised in Table~\ref{tab:fid} at the various orders in $\alphas$.
Note that in the case of the exclusive process, we employ Eq.~\eqref{eq:exerr} to estimate theory uncertainties but also provide the variant based on the correlated scale variation in parenthesis.

\begin{table}
	\centering
    \renewcommand{\arraystretch}{1.6}
    \begin{ruledtabular}
        \begin{tabular}{rcc}
            &
            $\PWp\PH$+$\geq1$jet &
            $\PWp\PH$+1jet \\
            \colrule
            $\sigma^{\text{LO}} \, [\si{\fb}]$  &
            $\num{20.99}\,^{+\num{2.09}}_{-\num{1.83}}$ &
            $\num{20.99}\,\pm\num{1.96}\,\bigl({}^{+\num{2.09}}_{-\num{1.83}}\bigr)$ \\
            $\sigma^{\text{NLO}} \, [\si{\fb}]$ &
            $\num{26.12}\,^{+\num{0.94}}_{-\num{0.99}}$ &
            $\num{17.42}\,\pm\num{2.10}\,\bigl({}^{+\num{0.73}}_{-\num{1.35}}\bigr)$ \\
            $\sigma^{\text{NNLO}}\, [\si{\fb}]$ &
            $\num{26.36}\,^{+\num{0.04}}_{-\num{0.24}}$ &
            $\num{15.59}\,\pm\num{0.59}\,\bigl({}^{+\num{0.48}}_{-\num{0.44}}\bigr)$ \\
        \end{tabular}
    \end{ruledtabular}
    \caption{The fiducial cross sections for the experimental set-up at \SI{13}{\TeV} detailed in the main text.
        The lower and upper error estimates are obtained from the envelope of the cross section values evaluated at the seven different scales.
        For the exclusive 1-jet predictions, the (symmetric) errors are obtained from the uncertainties of the inclusive 1-jet and 2-jet predictions, added in quadrature.}
	\label{tab:fid}
\end{table}

In the case of the inclusive production, we observe a drastic stabilization of the fiducial cross section at NNLO: the correction to the NLO central value
is of~$\order{1\%}$ and the theoretical uncertainty reduces from~$\order{4\%}$ at NLO to less than $\order{1\%}$ at NNLO.
Furthermore, the NNLO value is fully contained within the scale uncertainty of the NLO prediction, indicating a stable perturbative convergence.

In contrast, for exclusive production, the higher-order corrections systematically suppress the cross section and the scale uncertainties are larger.
The uncertainty estimate at NLO is similar to that at LO, with a significant reduction only observed when going to NNLO for both the conservative uncorrelated scale variation (Eq.~\eqref{eq:exerr} ) and the correlated scale variation given in parenthesis. We note that only the more conservative variant is able to reliably estimate uncertainties with subsequent orders overlapping in their uncertainty intervals.

To study the numerical impact of the top-loop induced parts, we compare the coefficient of the fiducial cross section obtained at $\order*{\alphas^2y_t}$ with the inclusive/exclusive NNLO Drell--Yan-like coefficients at $\order*{\alphas^3}$ computed at the central scale $\muF =\muR= M_{\PW\PH}$:
\begin{align*}
    \delta \sigma({\alphas^2 y_t}) &= 0.32\,^{+0.07}_{-0.06} \,\si{\fb}, \\[2ex]
    \delta \sigma_{\geq1\text{j}}({\alphas^3}) &= 0.24 \,\si{\fb}, &
    \delta \sigma_{1\text{j}}({\alphas^3}) &= -1.83 \,\si{\fb}.
\end{align*}
We observe that the $\order*{\alphas^2 y_t}$ top Yukawa-induced piece is of the same order of magnitude as the inclusive Drell--Yan-like $\order*{\alphas^3}$ correction and much smaller than the exclusive one.
The size of this top loop-induced piece is comparable to the uncertainty on both the inclusive and exclusive NNLO Drell--Yan-like cross sections (cf.~Table~\ref{tab:fid}), which necessarily prompts its inclusion for precision phenomenology.
However, the theoretical error estimate on these top-loop contributions is small, and as such, we do not expect their---mostly unknown---higher-order $\order*{\alphas^3 y_t}$ corrections to be phenomenologically relevant for this process.

%

\subsection{Distributions}

Moving on to kinematic distributions, we will present the results in figures that are divided into four panels.
The top panel shows the absolute predictions; the two centre panels show the $K$-factors for the inclusive and exclusive process, respectively. For the latter we further include the error bands based on a correlated scale variation shown as shaded bands.
Finally, the bottom panel shows the veto efficiency defined as
\begin{align*}
    \epsilon_\text{veto}(\mathcal{O}) =
    \frac{\mathrm{d}\sigma_{1\text{j}}     / \mathrm{d}\mathcal{O}}
         {\mathrm{d}\sigma_{\geq1\text{j}} / \mathrm{d}\mathcal{O}}
         ,
\end{align*}
for an observable $\mathcal{O}$.

\begin{figure}
    \includegraphics[width=0.95\columnwidth]{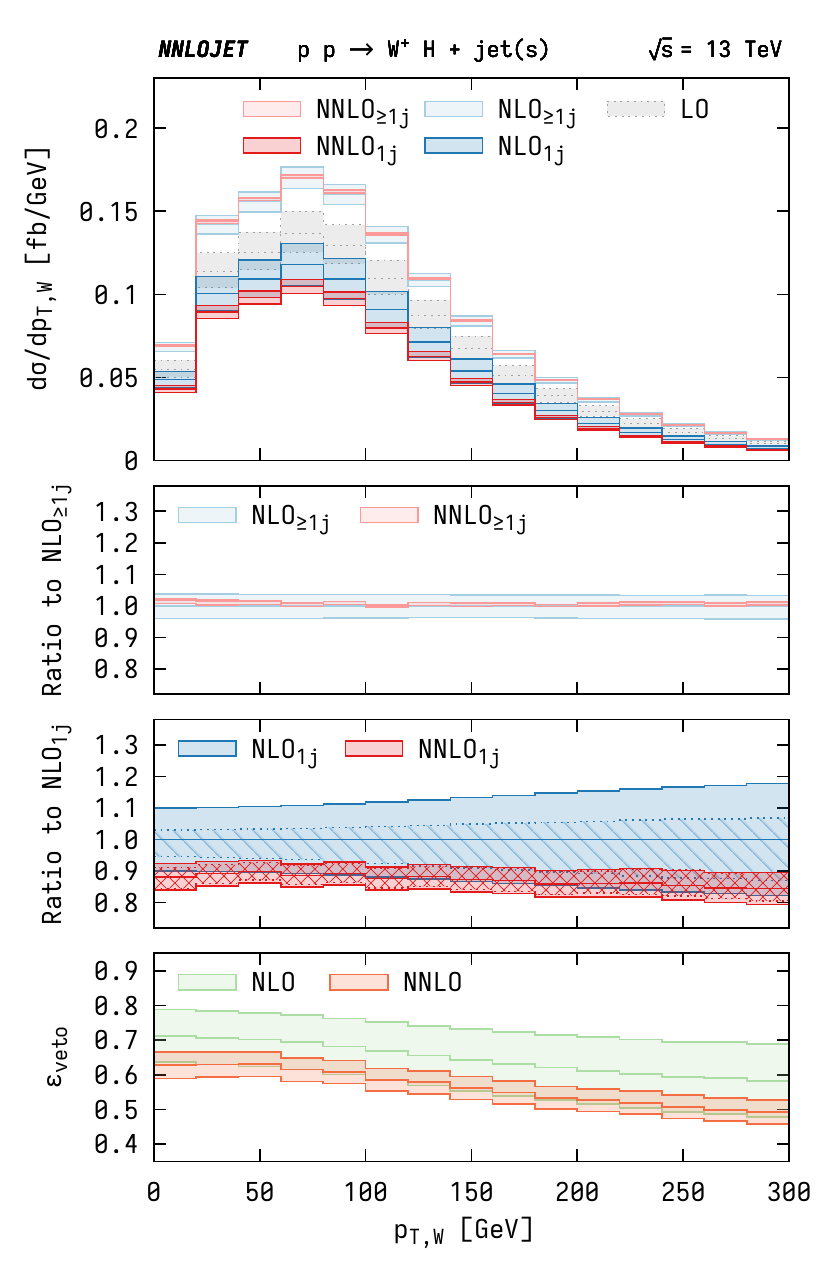}
    \caption{$W$ transverse momentum distribution for the $\PWp\PH$ +jet production process.
    The panels are described in the main text.}
    \label{fig:wphj_ptw}
\end{figure}

Fig.~\ref{fig:wphj_ptw} shows the $\PWp$ boson transverse momentum distribution.
In both the inclusive and the exclusive case, the NNLO $K$-factors are found to be rather flat, and in the inclusive case very close to unity.
Similar to the observations made for the fiducial cross sections, the uncorrelated error estimates are important in obtaining overlapping uncertainty estimates going from NLO to NNLO in the exclusive production process.
The veto-efficiency decreases towards larger transverse momenta, as harder interactions are probed that are more likely to be accompanied by additional resolved QCD emissions.
The shape of $\epsilon_\text{veto}$ is already well captured with one additional emission (NLO), while the NNLO corrections give an overall shift accompanied by a reduction of the residual uncertainties.
The results for the transverse momentum of the Higgs boson (not shown here) follow a qualitatively very similar pattern.

\begin{figure*}
    \includegraphics[width=0.95\columnwidth]{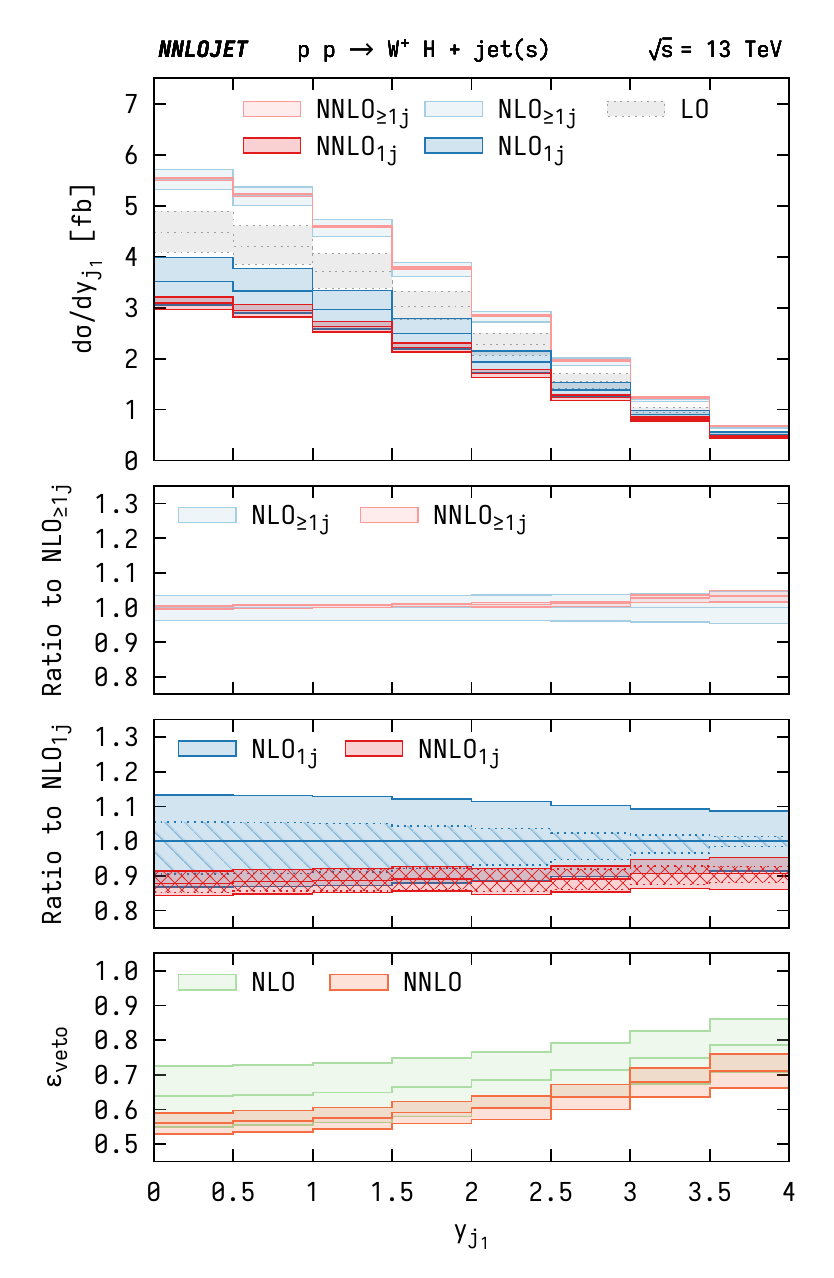}
    \includegraphics[width=0.95\columnwidth]{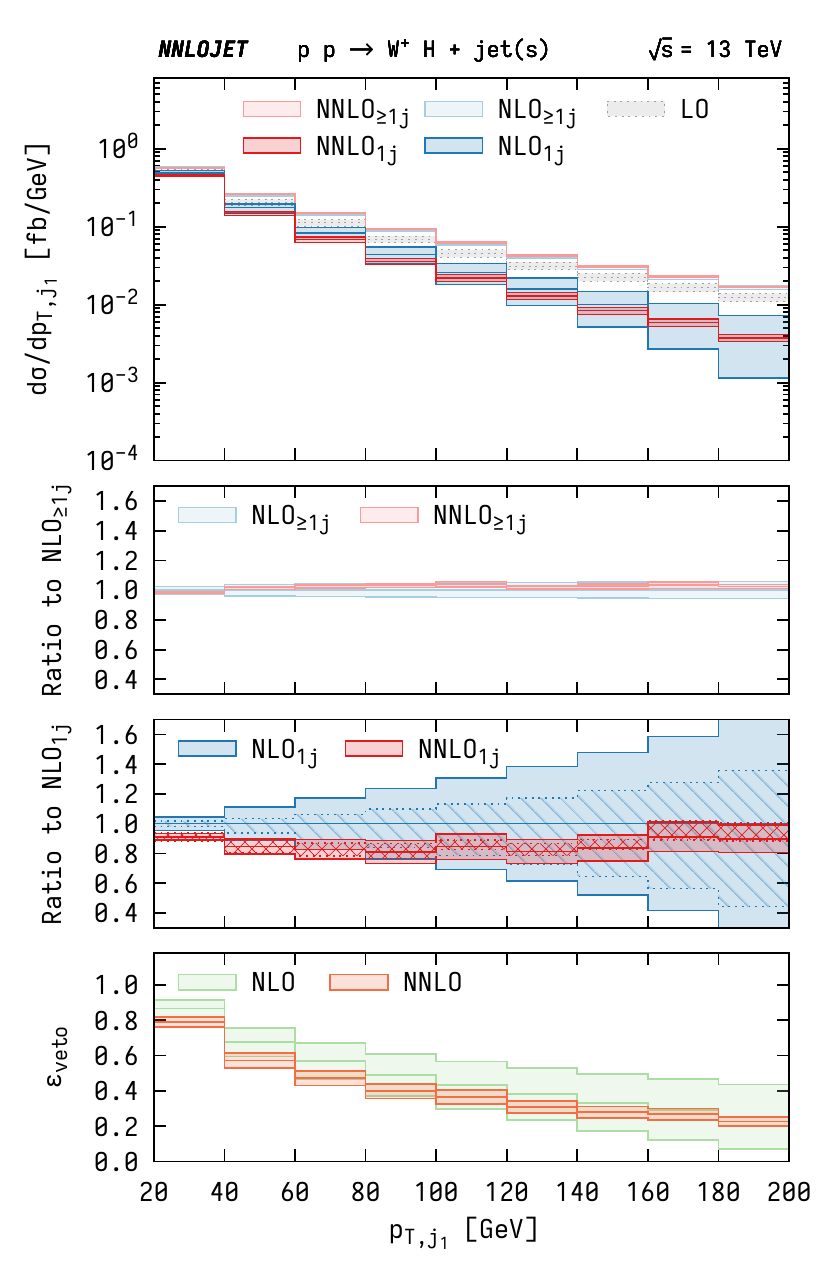}
    \caption{Leading jet rapidity~(left) and transverse momentum~(right) for the $\PWp\PH$ +jet production process.
    The panels are described in the main text.}
    \label{fig:wphj_j1}
\end{figure*}

Fig.~\ref{fig:wphj_j1} shows the rapidity (left) and the transverse momentum (right) of the leading jet.
The rapidity distribution, due to its inclusive nature in transverse momentum, follows a similar pattern to the fiducial cross section and the $p_{\mathrm{T},\PW}$ distribution discussed above: NNLO $K$-factors that are rather flat (with the exception of high rapidities in the inclusive case).
The veto efficiency increases slightly towards forward rapidities, indicating that a leading jet produced in the very forward region is less likely to be accompanied by additional hard resolved emissions.
The transverse momentum of the leading jet, on the other hand, shows very large corrections with large uncertainties in the high-$p_{\mathrm{T}}$ tail.
This can be explained by the fact that this region is dominated by two high-$p_\mathrm{T}$ jets recoiling against each other, while the colour-neutral $\PW\PH$ system is produced almost at rest.
As a consequence, the exclusive process is strongly suppressed in the tail and the formal accuracy of the prediction effectively degrades by an order.
The scale uncertainties of the NNLO prediction are therefore at the level of $\pm10\%$ here, which is more characteristic of an NLO prediction.
Further theoretical improvement in this kinematic regime could be achieved by considering jet-veto resummation in the presence of a hadronic jet~\cite{Liu:2012sz}.

Overall, we observe that the inclusive process exhibits an excellent perturbative convergence with small corrections and tiny residual scale uncertainties.
The observables in the exclusive process receive larger QCD corrections and the error prescription of Eq.~\eqref{eq:exerr} is crucial in obtaining overlapping uncertainty bands and thus reliable estimates for them.
The veto efficiencies are already well captured at NLO, with the NNLO prediction lying well within the uncertainty estimate of the previous order with uncertainty bands that are typically reduced by more than a factor of two.

\section{Conclusions and Outlook}


We have presented the computation of precise predictions for differential observables related to the associated production of an on-shell Higgs boson with a (leptonically decaying) charged vector boson and a hadronic jet for proton-proton collisions at \SI{13}{\TeV}.
These predictions include both Drell--Yan-like and top quark loop-induced contributions, for which we have included QCD corrections up to $\order*{\alphas^3}$ and $\order*{\alphas^2y_t}$ for the first time.

We have considered observables related to both inclusive and exclusive jet rates.
In the case of inclusive jet production, the perturbative corrections to the central value are small (flat K-factors close to unity) and the residual theoretical uncertainties are considerably reduced.
For exclusive jet production, the perturbative corrections are $\order*{-10\%}$ negative and the theoretical uncertainty is reduced to $\order*{5\%}$ for the considered distributions.
It is found that the NLO and NNLO predictions for the exclusive process are consistent only when the uncorrelated prescription for evaluating the theoretical uncertainty in exclusive jet rates is applied.
This is an important result as it verifies that the current approach taken by the experimental collaborations~\cite{Aad:2020jym,Aad:2020eiv} to evaluate the theoretical uncertainty on the signal process is reliable.

The theoretical modelling of the signal process, defined in terms of exclusive jet bins, contributes to one of the main sources of systematic uncertainty in the experimental measurements of the $V\PH$(+jet) process, and we have shown here how this uncertainty can be substantially reduced through the inclusion of NNLO QCD corrections.
In the future, the computation of all Higgs Strahlung modes (including a negatively charged or a neutral gauge boson in association with a hard jet) will allow for a comprehensive study of the theoretical uncertainties for all $V\PH$(+jet) modes with high precision.
Such a study will be vital in reducing the uncertainty associated to the signal modelling in future $V\PH$(+jet) measurements at the LHC, which will ultimately improve the experimental sensitivity to the Higgs-boson couplings.
Such a study is envisaged for future work.

\begin{acknowledgments}
We would like to thank Jonas Lindert for facilitating the use and inclusion of OpenLoops amplitudes into our computations, and to Hannah Arnold, Brian Moser, Tristan du Pree for discussions on experimental aspects of this work.
Furthermore we thank Xuan Chen, Juan Cruz-Martinez, James Currie, Thomas Gehrmann, Marius Höfer, Tom Morgan, Jan Niehues, João Pires, Duncan Walker, and James Whitehead for useful discussions and their many contributions to the \nnlojet code.
This research is supported by the Dutch Organisation for Scientific Research (NWO) through the VENI grant 680-47-461 and by the Swiss National Science Foundation (SNF) under contract 200021-172478.
\end{acknowledgments}

\bibliography{whj}

\end{document}